\documentclass[a4paper,11pt]{article}
\usepackage{pos}
\usepackage[T1]{fontenc} 
\usepackage{subfig}
\usepackage{float}
\usepackage{xcolor}
\usepackage{adjustbox}
\usepackage{blindtext}
\usepackage{multirow}
\usepackage{verbatim}
\usepackage{bm}
\usepackage{mathrsfs}
\usepackage{amsfonts}
\usepackage{tikz-feynman}
\usepackage{amsmath}
\usepackage{slashed}
\usepackage[normalem]{ulem}
\usepackage{cleveref}
\usepackage{feynmf}
\usepackage{multicol}
\usepackage{booktabs}
\usepackage{contour}
\usepackage{url}
\definecolor{Gray}{gray}{0.9}
\newcommand{\beq}{\begin{equation}}
\newcommand{\eeq}{\end{equation}}
\newcommand{\bea}{\begin{eqnarray}}
\newcommand{\eea}{\end{eqnarray}}

\renewcommand{\phi}{\ensuremath{\varphi}}
\newcommand{\sss}{\scriptscriptstyle}

\newcommand{\OO}{\mathcal{O}}

\newcommand{\Opp}[2]{\OO_{\sss #1}^{\sss #2}}

\newcommand{\tttt}{$t\bar{t}t\bar{t}$\xspace}
\DeclareRobustCommand{\rchi}{{\mathpalette\irchi\relax}}
\newcommand{\irchi}[2]{\raisebox{\depth}{$#1\chi$}}

\title{Complete study of four top quark production at the LHC in the SMEFT}

\author{Hesham El Faham}

\affiliation{Centre for Cosmology, Particle Physics and Phenomenology (CP3),\\ Universit\'e Catholique de Louvain,\\ Chemin du Cyclotron, B-1348 Louvain la Neuve, Belgium}
\affiliation{Inter-University Institute for High Energies (IIHE), Vrije Universiteit Brussel, \\Pleinlaan 2, 1050 Brussels, Belgium}

\emailAdd{hesham.el.faham@vub.be}

\abstract{We present a study of four top quark production at the LHC in the Standard Model Effective Field Theory (SMEFT). 
The analysis is performed at the tree-level, including all terms from the mixed QCD-EW-coupling cross-section expansion and all the relevant dimension-six SMEFT operators.  
We find several cases in which formally subleading terms give rise to significant contributions.
Moreover, we obtain limits on the SMEFT Wilson coefficients (WCs), including different collider setups, FCC-hh and HL-LHC, using a simplified chi-square fit. Through the latter, we also assessed the importance of including differential information in the fit.}

\FullConference{%
  41st International Conference on High Energy physics - ICHEP2022\\
  6-13 July, 2022\\
  Bologna, Italy
}

\begin{document}
\maketitle

\section{Introduction}
The four top quark production at the LHC is characterised by a tiny SM cross-section at 13 TeV of about 12 fb.  
Despite this, \tttt signatures are distinctive, leading to an affluent and energetic final state rendering four top signatures a remarkable opportunity for probing new physics.

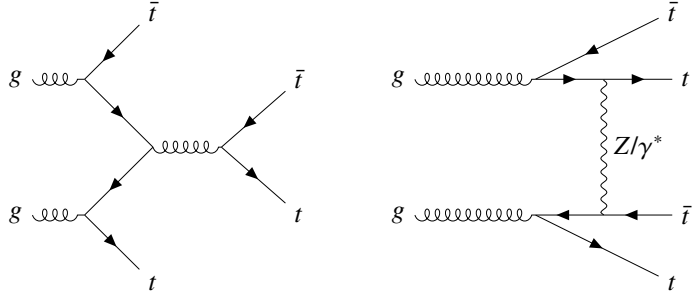
\begin{figure}[h!]
    \centering
    \scalebox{0.9}{\begin{tikzpicture}
\begin{feynman}[small]
\vertex (a0) {};
\vertex[right = of a0] (a1) {};
\vertex[right = of a1] (a2) {$\bar{t}$};
\vertex[below = of a0] (c0) {$g$};
\vertex[right = of c0] (c1);
\vertex[right = of c1] (c2) {};
\vertex[right = of c2] (c3) {};
\vertex[right = of c3] (c4) {$\bar{t}$};
\vertex[right = of c4] (c5) {};
\vertex[below = of c0] (b0) {};
\vertex[right = of b0] (b1) {};
\vertex[right = of b1] (b2);
\vertex[right = of b2] (b3);
\vertex[right = of b3] (b4) {};
\vertex[right = of b4] (b5) {};
\vertex[below = of b0] (d0) {$g$};
\vertex[right = of d0] (d1);
\vertex[right = of d1] (d2) {};
\vertex[right = of d2] (d3) {};
\vertex[right = of d3] (d4) {$t$};
\vertex[right = of d4] (d5) {};
\vertex[below = of d0] (e0) {};
\vertex[right = of e0] (e1) {};
\vertex[right = of e1] (e2) {$t$};
\diagram*{
(c0) -- [gluon] (c1),
(d0) -- [gluon] (d1),
(c1) -- [fermion] (b2),
(c1) -- [anti fermion] (a2),
(d1) -- [anti fermion] (b2),
(d1) -- [fermion] (e2),
(b2) -- [gluon] (b3),
(b3) -- [anti fermion] (c4),
(b3) -- [fermion] (d4),
};
\end{feynman}
\end{tikzpicture}
\begin{tikzpicture}
\begin{feynman}[small]
\vertex (a0) {};
\vertex[right = of a0] (a1) {};
\vertex[right = of a1] (a2) {};
\vertex[right = of a2] (a3) {};
\vertex[right = of a3] (a4) {$\bar{t}$};
\vertex[below = of a0] (c0) {$g$};
\vertex[right = of c0] (c1) {};
\vertex[right = of c1] (c2);
\vertex[right = of c2] (c3) ;
\vertex[right = of c3] (c4) {$t$};
\vertex[right = of c4] (c5) {};
\vertex[below = of c0] (b0) {};
\vertex[right = of b0] (b1) {};
\vertex[right = of b1] (b2);
\vertex[right = of b2] (b3);
\vertex[right = of b3] (b4) {};
\vertex[right = of b4] (b5) {};
\vertex[below = of b0] (d0) {$g$};
\vertex[right = of d0] (d1){};
\vertex[right = of d1] (d2);
\vertex[right = of d2] (d3);
\vertex[right = of d3] (d4) {$\bar{t}$};
\vertex[right = of d4] (d5) {};
\vertex[below = of d0] (e0) {};
\vertex[right = of e0] (e1) {};
\vertex[right = of e1] (e2) {};
\vertex[right = of e2] (e3) {};
\vertex[right = of e3] (e4) {$t$};
\diagram*{
(c0) -- [gluon] (c2),
(d0) -- [gluon] (d2),
(c2) -- [anti fermion] (a4),
(d2) -- [fermion] (e4),
(c2) -- [fermion] (c3),
(d2) -- [anti fermion] (d3),
(c3) -- [boson, edge label = $Z$/$\gamma^{*}$] (d3),
(c3) -- [fermion] (c4),
(d3) -- [anti fermion] (d4),
};
\end{feynman}
\end{tikzpicture}}
    \caption{Representative diagrams of four top production at $\mathscr{O}(\alpha_{s}^{2})$ (\emph{left}) and at $\mathscr{O}(\alpha_{s}\alpha_{\mathrm{w}})$ (\emph{right}), at the LHC.}
    \label{fig:4tops_LO_diags}
\end{figure}

The complete NLO predictions, including all possible QCD and electroweak (EW) orders, have been calculated in Ref.~\cite{Frederix:2017wme}.  
The latter's results show a peculiar and unforeseen interplay between EW and QCD contributions, with significant contributions arising from formally subleading terms.  
In the SM, representative diagrams of purely QCD-induced four-top production and through the inclusion of EW vertices are shown in \cref{fig:4tops_LO_diags}.
QCD-induced diagrams typically provide the leading contribution. 
However, Ref.~\cite{Frederix:2017wme} shows that in the SM at $\sqrt{s}=13$ TeV, the contributions arising from the non-purely-QCD diagrams are significant, reaching more than a third of the leading tree-level QCD contributions. 
Such a peculiar cross-section behaviour motivates a detailed study of new physics effects, particularly in the SMEFT framework.

\section{Four-fermion operators and cross-section expansion}
This work considers all possible contributions from the SM and SMEFT, including all dimension-six operators of the latter~\cite{Aoude:2022deh}. 
We follow the notation and operator conventions of Refs.~\cite{Aguilar-Saavedra:2018ksv,Degrande:2020evl} and consider all possible dimension-six SMEFT operators contributing to four top quark production.
However, in this note, we mainly focus on the 4-heavy four-fermion operators (we refer the interested reader to Ref.~\cite{Aoude:2022deh} for the detailed study):
\begin{equation}%
    \{\Opp{QQ}{1},\Opp{QQ}{8},\Opp{tt}{1},\Opp{Qt}{1},\Opp{Qt}{8}\}.
    \label{eq:relevant_operators}
\end{equation}%
The decomposition of the partonic differential cross-section up to $\mathscr{O}(\Lambda^{-4})$ in SMEFT reads
\begin{align}%
    d\sigma = d\sigma_{\rm SM} +\frac{1}{\Lambda^2} d\sigma_{\rm int} + 
    \frac{1}{\Lambda^4}
    \big(
    d\sigma_{\rm quad} +  d\sigma_{\rm dbl} + d\sigma_{\rm d8}
    \big),
\end{align}%
where the leading SMEFT contribution, $d\sigma_{\rm int}$, is the linear interference between the SM amplitudes and ones with a single insertion of a dimension-six operator. The $d\sigma_{\rm quad}$ and $d\sigma_{\rm dbl}$ are the squared single-insertion and double-insertion contributions, respectively. 
The interference cross-section expansion for the $gg$-initiated four top quark production in the presence of four-fermion SMEFT operators can therefore be written as follows:
\begin{align}
    d\sigma_{{\rm int}, gg, \texttt{[4F]}}  
    &=\alpha_s^3\,d\sigma^{(3,0,0)}_{{\rm int},gg} 
    +\alpha_s^2\left(\,\alpha\,d\sigma^{(2,1,0)}_{{\rm int},gg} +\alpha_t\,d\sigma^{(2,0,1)}_{{\rm int},gg}\right),
    \label{eq:gg_xsec_4f}
\end{align}
with $\alpha_s$, $\alpha$ and $\alpha_{t}$ denoting the strong, gauge and top-Yukawa couplings, respectively.

\section{Inclusive predictions}
\label{sec:inclusive}
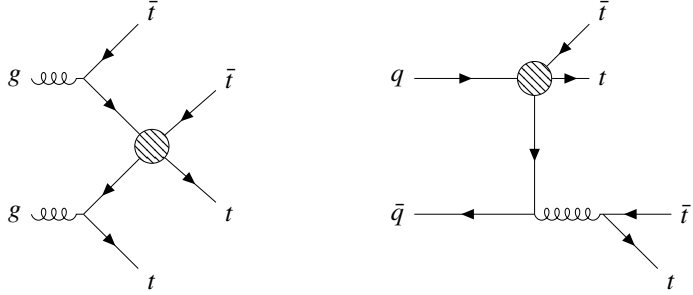
\begin{figure}[h!]
    \centering
    \scalebox{0.9}{\begin{tikzpicture}
\begin{feynman}[small]
\vertex (a0) {};
\vertex[right = of a0] (a1) {};
\vertex[right = of a1] (a2) {$\bar{t}$};
\vertex[below = of a0] (c0) {$g$};
\vertex[right = of c0] (c1);
\vertex[right = of c1] (c2) {};
\vertex[right = of c2] (c3) {$\bar{t}$};
\vertex[right = of c3] (c4) {};
\vertex[right = of c4] (c5) {};
\vertex[below = of c0] (b0) {};
\vertex[right = of b0] (b1) {};
\vertex [blob] (b2) at (2,-2) {};
\vertex[right = of b2] (b3);
\vertex[right = of b3] (b4) {};
\vertex[right = of b4] (b5) {};
\vertex[below = of b0] (d0) {$g$};
\vertex[right = of d0] (d1);
\vertex[right = of d1] (d2) {};
\vertex[right = of d2] (d3) {$t$};
\vertex[right = of d3] (d4) {};
\vertex[right = of d4] (d5) {};
\vertex[below = of d0] (e0) {};
\vertex[right = of e0] (e1) {};
\vertex[right = of e1] (e2) {$t$};
\diagram*{
(c0) -- [gluon] (c1),
(d0) -- [gluon] (d1),
(c1) -- [fermion] (b2),
(c1) -- [anti fermion] (a2),
(d1) -- [anti fermion] (b2),
(d1) -- [fermion] (e2),
(b2) -- [anti fermion] (c3),
(b2) -- [fermion] (d3),
};
\end{feynman}
\end{tikzpicture}
\begin{tikzpicture}
\begin{feynman}[small]
\vertex (a0) {};
\vertex[right = of a0] (a1) {};
\vertex[right = of a1] (a2) {};
\vertex[right = of a2] (a3) {$\bar{t}$};
\vertex[right = of a3] (a4) {};
\vertex[below = of a0] (c0) {$q$};
\vertex[right = of c0] (c1) {};
\vertex [blob] (c2) at (2,-1){};
\vertex[right = of c2] (c3) {$t$};
\vertex[right = of c3] (c4) {};
\vertex[right = of c4] (c5) {};
\vertex[below = of c0] (b0) {};
\vertex[right = of b0] (b1) {};
\vertex[right = of b1] (b2);
\vertex[right = of b2] (b3);
\vertex[right = of b3] (b4) {};
\vertex[right = of b4] (b5) {};
\vertex[below = of b0] (d0) {$\bar{q}$};
\vertex[right = of d0] (d1){};
\vertex[right = of d1] (d2);
\vertex[right = of d2] (d3);
\vertex[right = of d3] (d4) {$\bar{t}$};
\vertex[right = of d4] (d5) {};
\vertex[below = of d0] (e0) {};
\vertex[right = of e0] (e1) {};
\vertex[right = of e1] (e2) {};
\vertex[right = of e2] (e3) {};
\vertex[right = of e3] (e4) {$t$};
\diagram*{
(c0) -- [fermion] (c2),
(d0) -- [anti fermion] (d2),
(c2) -- [fermion] (d2),
(d2) -- [gluon] (d3),
(c2) -- [anti fermion] (a3),
(c2) -- [fermion] (c3),
(d3) -- [anti fermion] (d4),
(d3) -- [fermion] (e4),
};
\end{feynman}
\end{tikzpicture}}
    \caption{Representative diagrams of four top quark production with blobs representing the one dimension-six EFT insertion, in the $gg$-initiated production mode (\emph{left}) and in the $q\bar{q}$-initiated production mode (\emph{right}).}
    \label{fig:diags_EFT_4F}
\end{figure}

This section presents the complete LO SMEFT inclusive predictions for \tttt production at $\sqrt{s}=13$ TeV at the LHC. Predictions are shown for the 4-heavy four-fermion operators, ones connecting four heavy quark lines, as displayed in the \emph{left} diagram of \cref{fig:diags_EFT_4F} where the shaded blob depicts the SMEFT insertion.
For more results on the 2-heavy 2-light four-fermion operators (depicted in the \emph{right} diagram of \cref{fig:diags_EFT_4F}), as well as the two-fermion and purely bosonic operators, we remind the reader to consult Ref.~\cite{Aoude:2022deh}, where also the differential predictions are presented.

\begin{figure}[h!]
    \centering
    \includegraphics[width=0.5\textwidth]{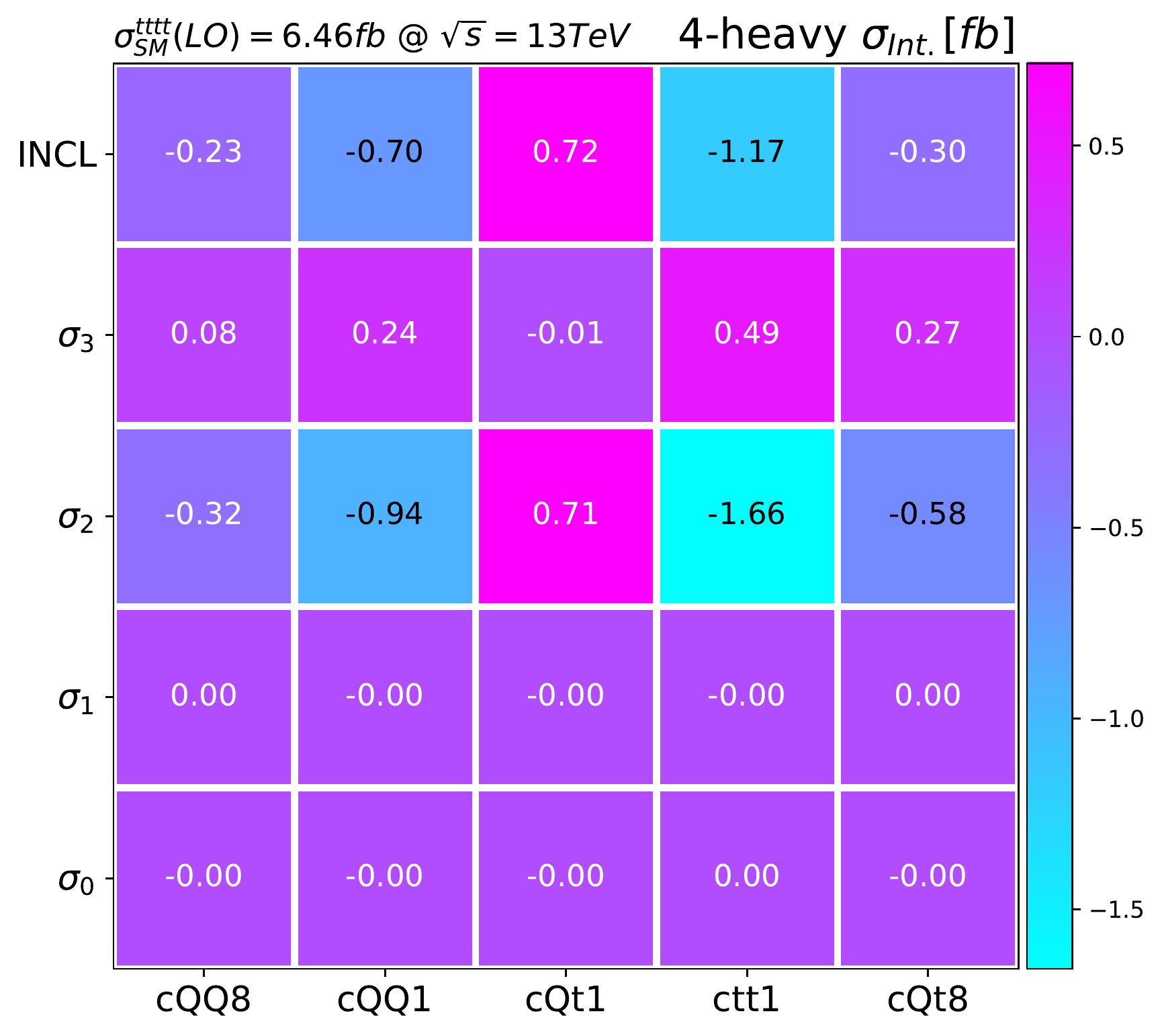}
    \caption{\label{fig:dim64f_4heavy_map_13TeV}
    Depiction of the interference strength of the 4-heavy operators.
    The columns denote the WCs in their UFO notation.
    The SM LO cross-section at $\sqrt{s}=13$ TeV is presented for reference.
    WCs were individually set to unity, and the scale of new physics $\Lambda$ is fixed to 1 TeV.}
\end{figure}

The computations were performed via \texttt{MadGraph5\_aMC@NLO}~\cite{Alwall:2014hca,Frederix:2018nkq} with the use of the \texttt{SMEFTatNLO} model~\cite{Degrande:2020evl}.
The interference cross-section of the 4-heavy four-fermion operators is presented in \cref{fig:dim64f_4heavy_map_13TeV} where the total inclusive interference cross-section, $\sigma_{\rm incl.}$/\texttt{INCL}, is split as follows: $\sigma_{\rm incl.}=\sigma_{3}+\sigma_{2}+\sigma_{1}+\sigma_{0}$,
with $\sigma_{i}$ with $i=3,2,1,0$ denoting the different contributions arising from terms with order $\alpha_{s}^{i}$ in line with \cref{eq:gg_xsec_4f}.
For \emph{all} 4-heavy operators, the dominant interference arises from formally subleading orders, $\sigma_{2}$.
This observation conflicts with the ``naive'' expectation that purely QCD-induced terms deliver the highest contribution to the cross-section (through $\sigma_{3}$). Such conflict emphasises the significance of the EW $tt \to tt$ scattering present in \tttt production at LO.

\section{Toy fits} 
\label{sec:toy}
This section presents limits on the WCs from a simplified $\rchi^{2}$ individual fit in different collider scenarios: LHC, FCC-hh and HL-LHC. Through such exercise, we explored the effects of differential information and the different collider energies on the WCs bounds. 

\begin{figure}[h!]
    \centering
    \includegraphics[width=0.6\textwidth]{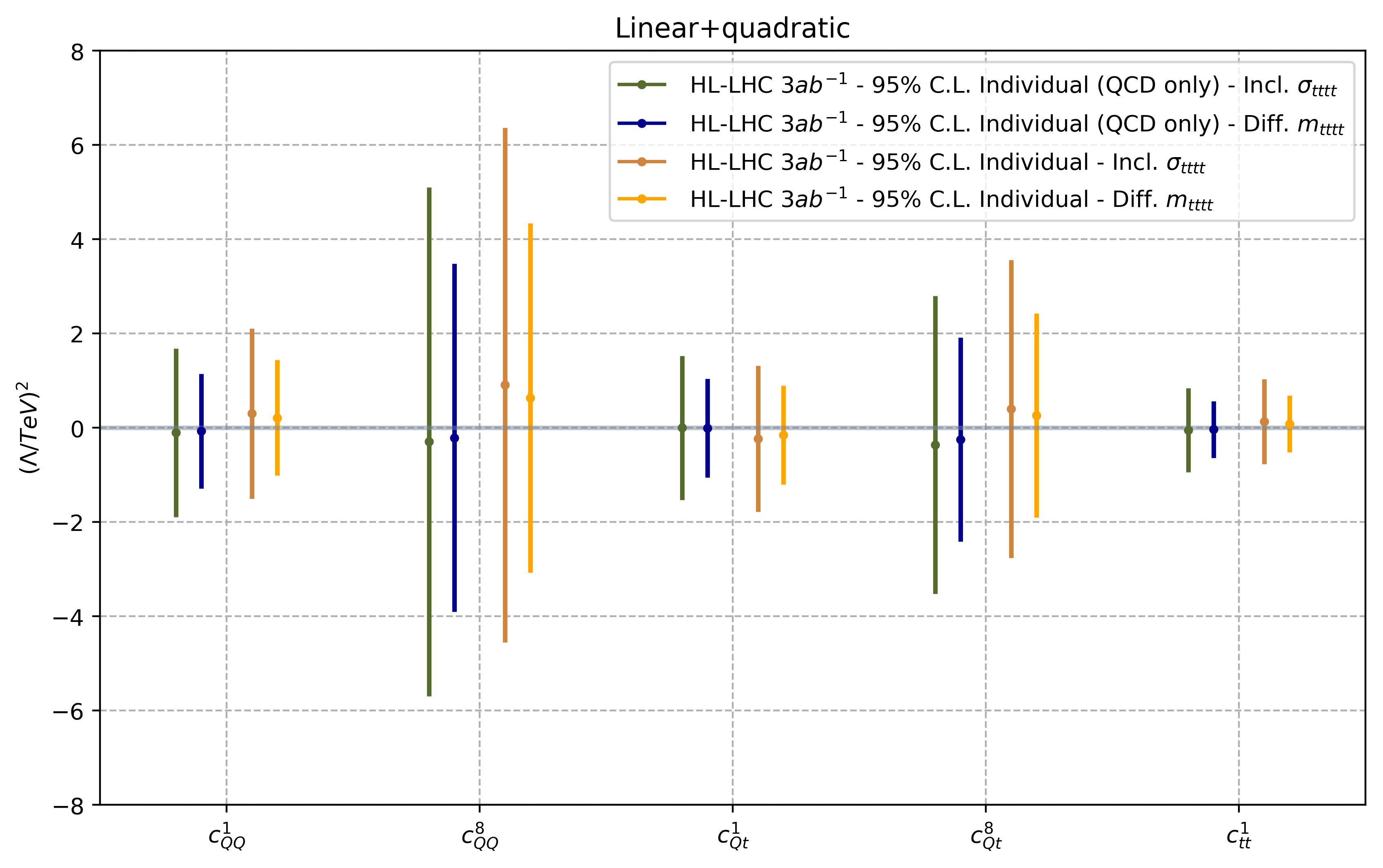}
    \caption{95\%CL limits on the 4-heavy operators' coefficients at the HL-LHC scenario from a $\rchi^{2}$ fit.
    The limits are shown for when using inclusive and differential information from $\sigma_{tttt}$ and $m_{tttt}$, respectively.}
    \label{fig:fit_results_bar_plot_HLHC}
\end{figure}

\paragraph{Impact of differential information} 
\Cref{fig:fit_results_bar_plot_HLHC} displays the individual limits presented (I) when only QCD-induced (leading) terms are considered and (II) when contributions to the cross-section from all tree-level terms in the mixed QCD-EW expansion are included. 
Furthermore, it compares the use of only inclusive information from $\sigma_{tttt}$ to when also adding differential information in the fit from $m_{tttt}$.
The exercise is performed for the HL-LHC at $\sqrt{s}=14$ TeV with 3 $\mathrm{ab}^{-1}$ of integrated luminosity.
The SM prediction calculated at LO is $\sigma_{tttt}^{\rm HL}=9.0$ fb, with a 20\% theoretical uncertainty.
Experimental measurement is assumed to be that of SM within the expected 28\% experimental total uncertainty~\cite{Azzi:2019yne}.
Our results indicate that differential information improves the sensitivity and should be used whenever possible. 

\begin{figure}[h!]
    \centering
    \includegraphics[width=1.0\textwidth]{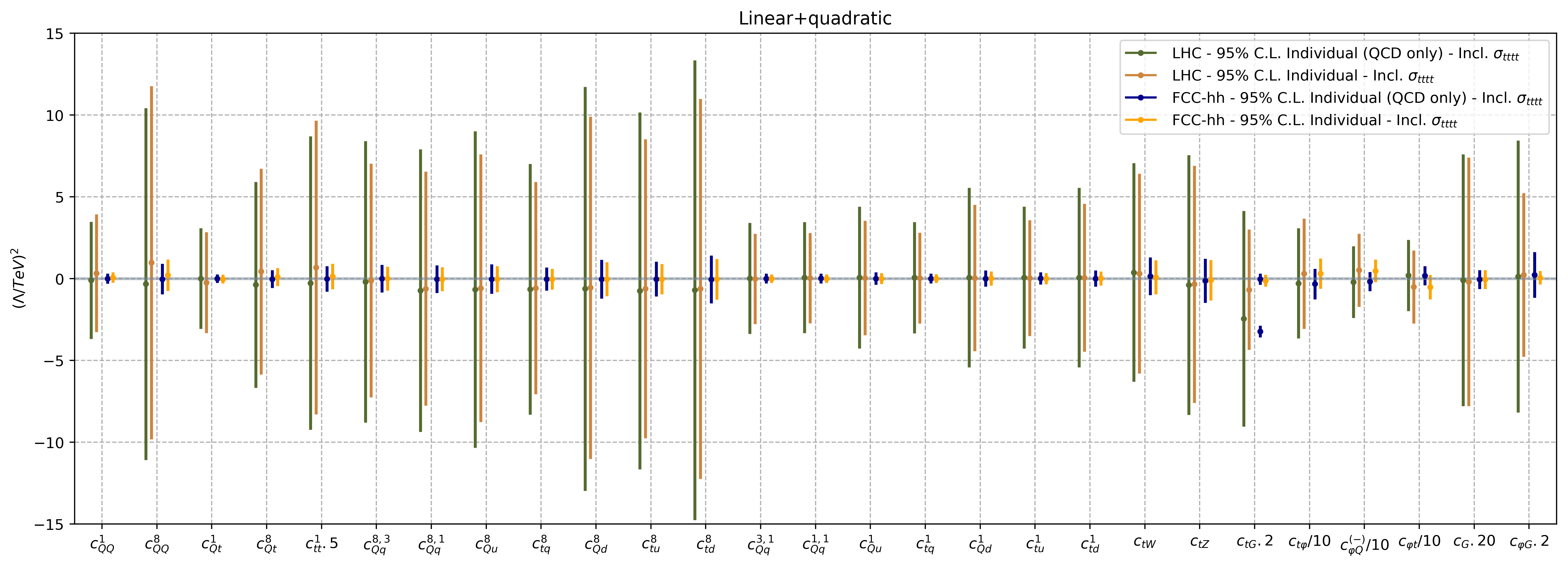}
    \caption{Limits on all the relevant operators used in the full study of Ref.~\cite{Aoude:2022deh} obtained from the $\rchi^{2}$ fit.}
    \label{fig:fit_results_bar_plot}
\end{figure}

\paragraph{Comparison of different collider setups} 
Finally, in \cref{fig:fit_results_bar_plot}, we compare the results from current LHC measurements to the FCC-hh bounds when only using inclusive information.
For the LHC, we use the SM prediction at NLO in QCD of Ref.~\cite{Frederix:2017wme}, and we fit the theoretical predictions to the inclusive ATLAS~\cite{ATLAS:2020hpj} and CMS~\cite{CMS:2019rvj} measurements.
The results from this fit show the significant constraining power the FCC-hh will be capable of providing on the SMEFT.

\section{Summary and conclusions}
We emphasise the importance of our results in summarising the four top quark production SMEFT predictions and analysing them in each order of the QCD-EW expansion.
Our observations underline the significance of subleading terms for the 4-heavy quark operators.
Moreover, within all the dimension-six SMEFT operators, we have shown that four top quark production provides strong constraints on the 4-heavy coefficients. We also showed that including differential information further improves the bounds on the SMEFT coefficients.

\section*{Acknowledgments}
I thank the organisers of ICHEP2022 for the invitation to share my work and for delivering a successful conference.
This work was supported from the European Union's Horizon 2020 research and innovation program as part of the Marie Skłodowska-Curie Innovative Training Network MCnetITN3 (grant agreement no. 722104), the F.R.S.-FNRS under the ``Excellence of Science'' EOS be.h project numbers 30820817 and 40005600, the European Research Council (ERC) under the European Union's Horizon 2020 research and innovation programme (grant agreement No. 949451), and from a Royal Society University Research Fellowship through grant URF/R1/201553.

\bibliographystyle{JHEP}
\bibliography{PosBib}

\end{document}